\documentclass[sigconf, 10pt, nonacm]{acmart}
\title{A Nationwide Snapshot of Starlink Affordability}

\author{Shaddi Hasan}
\affiliation{%
  \institution{Virginia Tech}
  \city{Blacksburg}
  \state{Virginia}
  \country{USA}
}
\email{shaddi@vt.edu}

\begin{abstract}
Starlink positions itself as ``fast affordable Internet'' that can provide service to nearly every household in the United States, and is slated to be responsible for providing the only broadband service to a significant number of US households via the NTIA's BEAD program.
However, neither the FCC's National Broadband Map nor Starlink's own availability map captures what service costs.
We present the first national census of Starlink's residential broadband pricing.
We first identify that Starlink's offers are uniform within H3 resolution 6 grid cells, allowing us to collect plans and prices for every populated cell in the United States, more than 162,000 addresses, in October 2026.
We find that while Starlink's service prices are fairly consistent, the effective price of the most affordable plan varies across the country, driven by restricting sales of less expensive plans as well as ``demand surcharges'' ranging between \$100 and \$1,500.
Using a common affordability threshold of 2\% of monthly income, we find that service restrictions or surcharges make Starlink unaffordable for approximately 4.3 million households, and in total nearly a quarter of all US households (roughly 25.5 million) cannot afford Starlink. 
Finally, approximately 80\% of locations where Starlink is slated to receive BEAD funding face no constrained pricing today.

\end{abstract}

\usepackage{graphicx, xcolor, xspace, subcaption, hyperref}

\begin{document}

\maketitle
\pagestyle{plain}

\section{Introduction}
\label{s:intro}

Low earth orbit (LEO) satellite megaconstellations, such as Starlink, promise to serve both as a way to fill in gaps in terrestrial Internet coverage and to provide a competitive service option in areas that are already served, such as cities and suburban areas.
Despite the complexity of launching the large numbers of satellites required to maintain always-available service as well as of deploying ground stations globally to connect the constellation to the broader Internet, Starlink is proof-by-example that such systems can deliver residential broadband service at scale.

On the one hand, this represents an incredible opportunity to bridge the so-called digital divide.
Based on the Federal Communication Commission's (FCC) National Broadband Map~\cite{fcc_nbm}, Starlink is the only broadband\footnote{In this paper, we adopt the current FCC definition of broadband as Internet service providing at least 100Mbps download and 20Mbps upload speeds.} option today for nearly 3.8 million homes throughout the United States.
Starlink is slated to receive significant public subsidy funding via the BEAD program~\cite{bead,ntia2025beadrestructuring}, ostensibly to ensure that they can upgrade their network capacity to serve a portion of these millions of homes.

But is it enough to simply be covered by signal from Starlink's satellites?
Early conceptions of the ``digital divide'' focused on the gap between those who had access to Internet and those who did not (the ``haves'' and ``have nots''), framing the divide as a binary focused on whether network infrastructure was physically available at a location~\cite{ntia1995falling, ntia1998falling}.
This simple framing obscures the fact that the economic, social, educational, and health benefits of Internet access don't accrue to a household simply because radio waves hit the roof or a fiber optic cable is buried nearby: people must have the skills to appropriately leverage the Internet, devices with which to access it, and, most critically, the ability to afford the services offered to them.
These critiques inform modern definitions of ``digital inclusion'', with advocacy groups such as the National Digital Inclusion Alliance~\cite{ndia_digital_navigator_2024} describing it as ensuring access to ``affordable, robust broadband'' alongside devices and digital skills.
For Starlink, then, the question is not only whether its satellites can reach a household, but whether that household can afford the service once they do.

If Starlink's service isn't affordable, or if their network lacks capacity to serve parts of the country, then people who need it might be left behind, despite the FCC's National Broadband Map~\cite{fcc_nbm} showing that nearly the entire country has Starlink service today.
Unfortunately, Starlink's pricing and capacity information is not captured by the FCC's maps,\footnote{Prior to July 2026, Internet service providers like Starlink were required to make machine-readable "Broadband Consumer Labels"~\cite{fcc2022bcl} available via their websites. However, ISPs were not required to indicate where a particular service plan was available. The FCC rescinded this requirement in July 2026~\cite{fcc2026bcl}.}
leaving a significant gap in our understanding of where Starlink is realistically available today at affordable prices.

In this paper, we set out to address this gap by collecting what is to our knowledge the first national census of Starlink's residential broadband pricing.
We identify a spatial unit within which Starlink's plans and prices are consistently uniform -- at the H3~\cite{h3} resolution 6 cell, a hexagonal cell approximately 36~km\textsuperscript{2} -- and then proceed to collect service plans and pricing for all populated grid cells in the United States.
Our resulting dataset, collected in October 2026, comprises pricing and availability at more than 162,000 addresses and reflects pricing for every US state and the District of Columbia.

Our findings paint a picture of a network with consistent pricing across most of the United States, with a number of notable regional exceptions.
Starlink's effective lowest priced plan varies, driven by restricting where their lowest-cost plans are sold as well as adding "demand surcharges" ranging between \$100 and \$1,500.
Using a common affordability threshold for broadband as costing less than 2\% of a household's monthly income~\cite{broadbandcommission2025affordability,fcc2016affordability,pew2023broadbandmiddleclass}, we find that 25.5M US households -- nearly a quarter -- are not able to afford Starlink, and that Starlink's service restrictions or surcharges are responsible for making Starlink unaffordable for approximately 4.3 million additional households.
We further find that areas in which Starlink is in line to receive funding via the BEAD program via LEO Capacity Subgrants are largely in areas that do not reflect pricing constraints, though we note that these awards are not yet final as of this writing.

The remainder of this paper is organized as follows.
We first review related work (Section~\ref{s:related}).
Next, we outline our survey methodology, including our process for identifying Starlink's pricing granularity (Section~\ref{s:method}).
We then present our key survey results, covering both pricing and affordability (Section~\ref{s:results}).
Finally, we conclude in Section~\ref{s:conclusion} by discussing Starlink's recent shift toward dynamic pricing, and why ongoing, national-scale measurement of broadband pricing will become more important as prices change more frequently.

\section{Related work}
\label{s:related}

\textbf{Measuring broadband availability and pricing.}
Prior work on broadband availability and pricing leverages the Broadband Availability Tools (BATs) that ISPs use to expose service offerings to prospective customers.
Major et al.~\cite{nowans} queried BATs across nine major ISPs to map broadband coverage in the US.
BQT~\cite{bqt} (and later BQT+~\cite{koduru2026bqtplus}) extended this approach to collect data on plans and prices offered as well as availability, as well as to evaluate policy, such as the efficacy the Connect America Fund program~\cite{manda2024caf}.
Investigative journalists have used identical methods to explore ``digital redlining'' in which ISPs offer slower services at the same prices in lower-income and historically redlined neighborhoods~\cite{markup2022dollars,markup2022methods}.
We use the same general methodology in this work, with two key changes. First, our crawler is synthesized by our agentic framework for measuring broadband pricing and availability, Broadband Surveyor.
Second, we focus on Starlink and the unique constraints that a LEO megaconstellation poses when used as an access network, which prior work has not evaluated.

\textbf{Starlink performance and capacity.}
A robust literature has measured Starlink's performance and how it varies, using measurements from individual user terminals~\cite{michel2022firstlook}, browser-based and crowdsourced speed tests~\cite{kassem2022browser,mohan2024starlink}, measurements of Starlink endpoints at global scale~\cite{izhikevich2024hitchhiking,tran2025outage}, controlled experiments with user terminals~\cite{tanveer2023constellations,kataria2026obstructed}, and models that predict performance globally~\cite{benghe2026horizon}.
A related literature has evaluated Starlink's capacity~\cite{woo2025starlink,xlab_starlink,chen2024physical,bhosale2025failover}.
Our work builds on this literature by examining how Starlink prices its service, and also provides a new signal exposed by Starlink to indicate potential limits to their network's capacity, using direct evidence of where Starlink's pricing reflects potential capacity constraints.

\section{Methodology}
\label{s:method}

Our approach has three steps: (1) we build a crawler that queries Starlink's availability tool for plans offered at an address, (2) we identify a spatial unit within which Starlink's offers are uniform, and (3) we query one address in every such unit nationwide.
This mirrors the approach taken in prior work~\cite{bqt,nowans} that queries a consumer-facing Broadband Availability Tool (BAT) on an ISP's website to obtain service plans, pricing, and speeds available at a specific postal address.

\subsection{Querying Starlink's Availability Tool}
\label{s:method:bat}
Our crawler is produced by an automated agentic pipeline we developed, Broadband Surveyor; we leave a full description of this system to future work.
This pipeline produces a deterministic crawler written in Python that runs queries against the Starlink BAT.
To ensure accuracy, prior to deployment, we manually obtain pricing and plans for a set of reference addresses, then validate that our generated crawler produces the same results as seen by a human on these addresses.
During the measurement campaign, we monitor crawled responses to ensure we receive valid responses that reflect live prices (e.g., not cached) for each queried location, and further verify that any results reported by our generated crawler are present in the network traffic exchanged by our crawler and the Starlink BAT.

For each address queried, we collect the full set of plans offered at a location, including speeds, prices, and fees for each offered plan, as well as any available FCC Broadband Consumer Labels (BCLs) to compare with prices listed via the BAT.
We also record the checkout (cart) page for each address, which applies any promotions Starlink offers; because the checkout page prices only the default plan, which was the cheapest plan in every cell, other plans are recorded at list price.
Each query also records a full log of all HTTP traffic to facilitate auditing responses.

\textbf{Crawling Ethics.}
To avoid disrupting Starlink's BAT tool, we strictly rate limit our measurement campaign to stay under 10 queries per minute, with queries run sequentially.
During the crawl, we track error rates and response times from the BAT for signs of overload; we encounter none.
Our system never enters any contact information in any forms that it encounters to avoid creating false or misleading sales leads.
We argue that the public benefit of this affordability dataset outweighs the minimal additional load placed on the ISP's public-facing search tool.

\subsection{Starlink's Pricing Granularity}
\label{s:method:unit}

In principle, an ISP could offer different prices to every address it serves.
Prior work has shown that ISPs sometimes employ fine-grained pricing variations, resulting at times in digital redlining~\cite{markup2022dollars,bqt}.
However, the nature of satellite service makes such fine-grained pricing unlikely to be the case for Starlink, given capacity is uniformly available within a satellite's beam and that customer premise equipment is not tied to a specific address.
Thus, we hypothesize that there exists some spatial unit at which all residential locations will receive uniform pricing; if this hypothesis were true, we could complete a census of Starlink pricing by querying just one location in each such spatial unit.

\begin{figure}[t]
  \centering
  \includegraphics[width=\columnwidth]{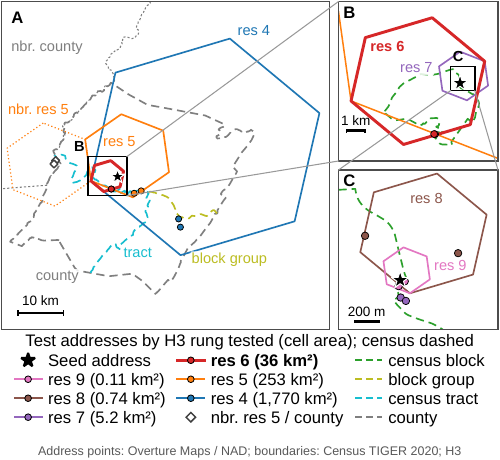}
  \caption{An example hierarchy of spatial units. Starting from the seed address, we select two addresses that fall inside each tier of the hierarchy but outside the next one.}
  \label{fig:ladder}
\end{figure}

As an initial exploratory investigation, we randomly select 158 seed locations across the country, and around each seed we build a hierarchy of possible spatial units: H3~\cite{h3} hexagonal grid cells from resolution 9 (about 0.1~km\textsuperscript{2}) up to resolution 4 (about 1,770~km\textsuperscript{2}), and administrative units including Census block, Census block group, Census tract, and county.
At each tier we pick two random addresses inside that tier but outside the one below it.
The smallest tier where answers start to differ from the seed's suggests a boundary where Starlink's offer stops being uniform.
We depict an example of such a spatial hierarchy around a single seed location in Figure~\ref{fig:ladder}.

We find that addresses in the same H3 resolution 6 cell (about 36~km\textsuperscript{2}) always receive the same answer, and prices only differ for locations in different resolution 6 cells.
Similarly, we see differences within administrative units, even at the Census block granularity, but only for addresses in different H3 resolution 6 cells.
Of the 2,066 addresses we test around our 158 seed addresses, 28 receive a different offer than their seed. 
These differences occur even between addresses in the same Census block, tract, or county, and in the same H3 resolution 4 or 5 cell, ruling out each of those as Starlink's pricing unit. However, every one of the 28 lies in a different H3 resolution 6 cell than its seed, and none of the 1,111 addresses inside the seed's own resolution 6 cell differs.

\textbf{Confirming the H3 Resolution 6 granularity.}
This initial experiment is suggestive, but as a sparse sample it does not confirm the granularity question -- most regions have uniform service offerings, making it rare to find variation.
After completing our national survey at the H3 resolution 6 level (Section~\ref{s:results}), we conduct a separate, targeted verification specifically along the borders of areas where we find differences in service plans.
Our goal is to validate that pricing does not vary at a finer spatial resolution; this ensures our survey does not miss price variation within resolution 6 cells.

We randomly select 25 boundaries between neighboring resolution 6 cells where our measurement observes differences in offers, excluding boundaries within 5~km of an international border and boundaries whose offers no longer differ at the time of this measurement.
At each boundary, we pick H3 cells at resolutions 7--10 (from about 0.015 to 5~km\textsuperscript{2}) that straddle the boundary, and then place one point on each side of the boundary within each straddling hexagon, plus a point at the center of each resolution 6 cell, a total of 1,650 points across 800 straddling hexagons.
Each point thus belongs to two distinct cells -- one at resolution 6 and one at the target resolution.
For every point, the price we measure aligns with the resolution 6 cell it belongs to, not any finer spatial resolution.
Because the two points within each straddling hexagon share the same finer cell but lie in different resolution 6 cells, they would receive the same price if Starlink priced at the finer resolution; instead, in every case they receive different prices, each matching its own resolution 6 cell.
This result makes it highly unlikely that Starlink sets prices at a finer granularity.

These results held over the duration of our survey and are congruent with how Starlink presents their own availability~\cite{starlink_map}.
Given these results, we proceed under the assumption that Starlink pricing is uniform within an H3 resolution 6 cell; for the remainder of this paper, we say "H3 cell" to refer specifically to an H3 resolution 6 cell.

\subsection{Identifying an Address Panel}
\label{s:method:panel}

Once we have identified Starlink's unit of variation, we define a panel of locations to query across the United States, one per H3 cell, to determine all residential Starlink prices.
We pick a random address per cell drawing upon a variety of address sources (including the National Address Database~\cite{nad}, Overture Maps~\cite{overture}, OpenAddresses~\cite{openaddresses}, and the BSL Fabric).
We then query this pool using the crawler produced by Broadband Surveyor.
Our address panel consists of 162,027 locations in 162,027 H3 resolution 6 cells,
which represents every single H3 cell in which Starlink claims residential service in the National Broadband Map in the 50 states and Washington DC where there is at least one location with an address.\footnote{We note that Starlink claims service in 4,238 cells in which no BSL has a usable address; these are largely concentrated in Alaska. We recognize this as a limitation of our address-based method and hope to address this gap in future work.}

Our crawler records response categories of "served with prices", "served with no prices shown", "not available", "inconclusive", or "unknown".
Of more than 162,000 queries, none were categorized as an unknown response, two were inconclusive, and four served with no prices shown.
"Not available" accounted for 39 responses, and the remainder (99.97\%) of responses were served with pricing shown.

We query cells in a random order, allowing each day's results to be a random sample of the whole country (about 6.5\% of H3 cells per day), so that a partial panel still represents a nationally representative sample.
This allows us to compare results over time while maintaining a representative geographic sample.

\section{Results}
\label{s:results}
We conducted a national survey of Starlink prices at 162,027 addresses from October 11 to October 26, 2026.
From this survey, we identified three primary findings.
First, Starlink maintains a small menu of prices that vary regionally, with demand-based surcharges and elimination of low-cost plans being the primary ways in which offerings vary across the country; we refer to these as "constrained" prices.
Second, based on the 2\% of monthly income threshold to define broadband affordability, we estimate Starlink is not currently affordable for approximately 25.5 million US households; constrained pricing makes Starlink unaffordable for an additional 4.3 million households.
Finally, while Starlink has not yet certified that they are providing BEAD-supported service, should they do so today, approximately 80\% of locations where Starlink is slated to receive LEO Capacity Subgrants from the BEAD program do not face constrained pricing today, while approximately 24,000 (about 5\%) currently require surcharges that are disallowed under BEAD rules.
We discuss each of these in turn.

\subsection{Starlink Prices Vary Regionally}
\label{ss:regional_variation}
We find that Starlink generally offers three residential service plans nationwide --- Residential 100~Mbps at \$55/mo, Residential 200~Mbps at \$85/mo, and Residential Max at \$130/mo.
The \$55/mo plan's \$349 rental kit is fully discounted at checkout, so in practice all three plans come with a free rental kit (plus \$20 shipping).
There are only a handful of H3 cells where Starlink is not available, which seem to be clustered near radio astronomy sites.
For the vast majority of the country, that means that anyone (though perhaps not everyone~\cite{xlab_starlink,woo2025starlink}) who has a clear view of the sky can get broadband service for \$55/mo today.
We also note a handful of significant location-based discounts, titled ``SpaceX Neighbor Discount'' and ``Vermillion\footnote{\textit{sic}; this apparently refers to the Vermilion Parish in Louisiana.} Parish Area Discount,'' which reduce the monthly service price by 50\% (e.g., to \$27.50/mo for the \$55/mo plan) in 363 H3 cells: the SpaceX Neighbor discount in 231 cells in Texas, Mississippi, Tennessee, and Arkansas, and the Vermilion Parish discount in 132 cells in Louisiana, the planned location of SpaceX's new Starbase Louisiana site.

There are a number of regions, however, where Starlink restricts its offer in graduated steps, first offering only its Max plan at \$130/mo, and then adding a one-time ``demand surcharge'' to that plan, ranging from \$100 to as much as \$1,500.
We refer to H3 cells with either restriction (e.g., Max-only or a demand surcharge) as ``constrained,'' and H3 cells with all three plans and no surcharge ``unconstrained.''

\begin{figure}[t]
  \centering
  \includegraphics[width=\columnwidth]{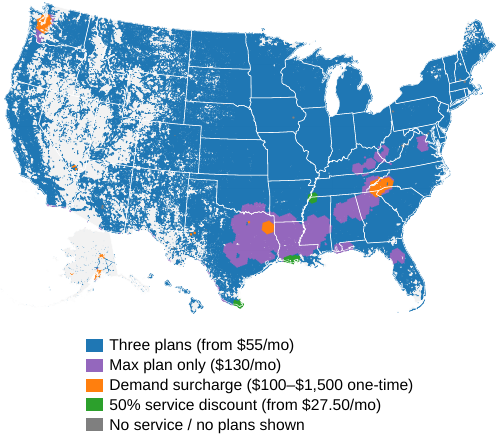}
  \caption{Purple and orange represent ``constrained'' areas with either only a single \$130/mo plan offered or a demand surcharge in place. Each point represents an H3 cell; only cells with residential households present are shown. Green marks cells with a 50\% location-based service discount, focused on SpaceX's Texas and Louisiana space launch facilities as well as around Memphis, TN, where SpaceX operates hyper-scale data centers.}
  \label{fig:map}
\end{figure}

These demand surcharges and limited plan offerings are always geographically clustered, as depicted in Figure~\ref{fig:map}.
The largest covers a swath from eastern Texas through Louisiana, southern Arkansas, Mississippi, Alabama, north Georgia, and east Tennessee, with demand surcharges around Asheville in western North Carolina and a \$500 surcharge zone on the Texas--Oklahoma border.
Other notable areas include eastern Kentucky and southern West Virginia, central Florida, the Puget Sound, a small area in northern Virginia, and the more densely populated areas of Alaska.

\begin{figure}[t]
  \centering
  \includegraphics[width=\columnwidth]{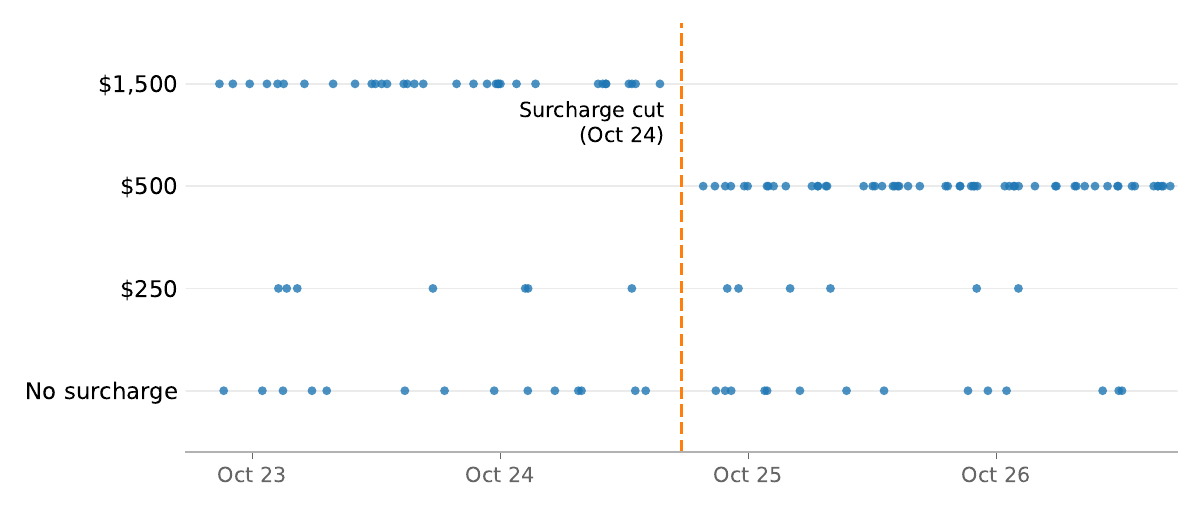}
  \caption{Each dot represents one query to one cell. After the cutoff on Oct 24, we see no more \$1,500 surcharges in Alaska.}
  \label{fig:alaska}
\end{figure}

\textbf{Offerings vary over time.} 
Because our survey samples a random set of cells every day, we can observe changes in what Starlink offers in an area soon after they occur.
For example, on October 24, 2026, we observed that Starlink cut Alaska's \$1,500 demand surcharge to \$500, allowing us to identify that the change took place within a roughly four-hour window on October 24 (Figure~\ref{fig:alaska}).
After this date, we observed no further \$1,500 surcharges in the state.
We did not observe any other such changes during our survey, however.

\begin{figure}[t]
  \centering
  \includegraphics[width=\columnwidth]{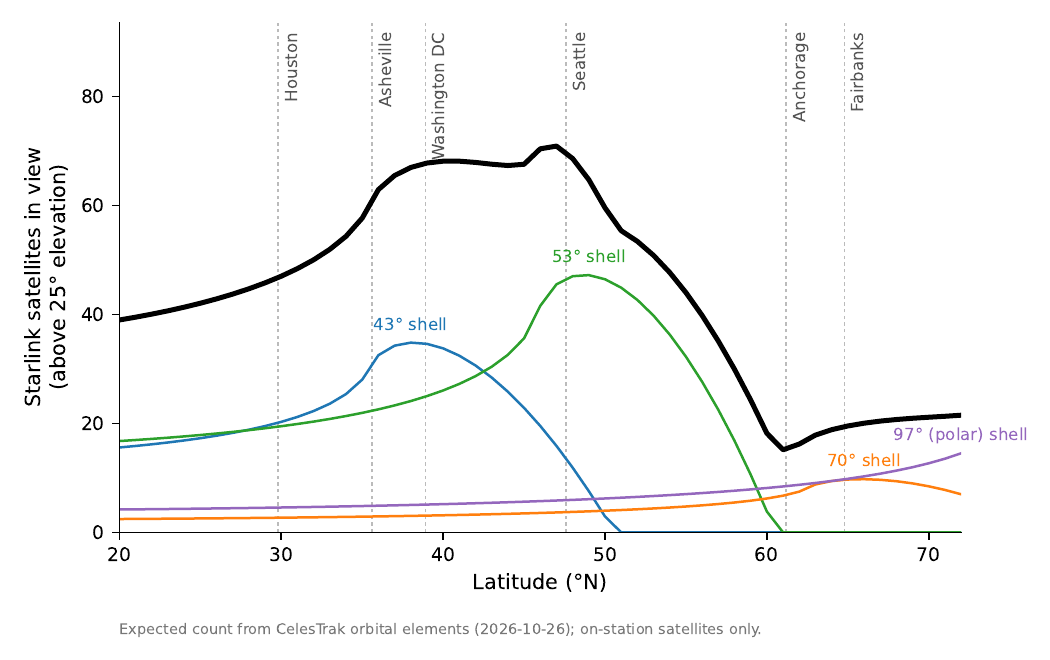}
  \caption{Starlink satellites visible (above 25\textdegree{} elevation) by latitude and shell, computed from CelesTrak orbital elements (2026-10-26). Number of visible satellites does not seem to obviously correspond to areas with constrained pricing.}
  \label{fig:latitude}
\end{figure}

\textbf{Causes for constrained prices are unclear.} We were not able to identify an obvious mechanism driving these pricing variations.
One simple explanation is a lack of satellites available to provide service, but this does not explain our results on its own, as shown in Figure~\ref{fig:latitude}.
Alaska relies mostly on Starlink's high-inclination and polar orbits
and sees about a quarter as many satellites as the mid-latitude US; Starlink has relatively few satellites visible from above about 50\textdegree N and especially above 60\textdegree N, meaning there are simply fewer satellites available to serve the state.
However, this does not explain constrained prices at other latitudes: Puget Sound, which sits at the best-covered latitudes in the country, also has surcharges.
Our prior work has suggested Starlink capacity could be limited by available spot beams~\cite{woo2026beyond}, but other factors could also drive resource scarcity, such as limited spectrum or limited ground stations.
There could also simply be higher demand for Starlink services in these areas, perhaps due to fewer affordable terrestrial options.
We leave identifying the underlying mechanisms to future work.

\subsection{Starlink Affordability}
\label{ss:afford}

\begin{figure*}[t]
  \centering
  \begin{subfigure}[t]{0.49\textwidth}
    \centering
    \includegraphics[width=\linewidth]{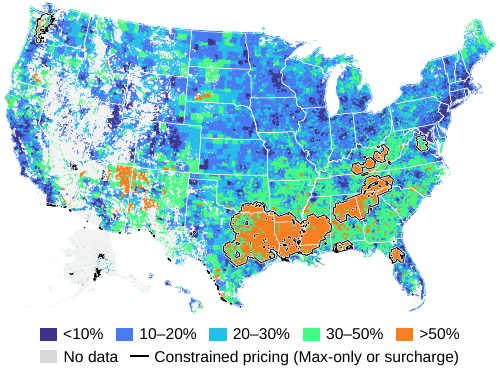}
    \caption{Estimated share of households in each H3 cell who cannot afford Starlink's lowest-cost offer (first-year cost including kit, surcharge, and shipping, at 2\% of income). Areas where Starlink offers constrained pricing are outlined.}
    \label{fig:afford_share}
  \end{subfigure}\hfill
  \begin{subfigure}[t]{0.49\textwidth}
    \centering
    \includegraphics[width=\linewidth]{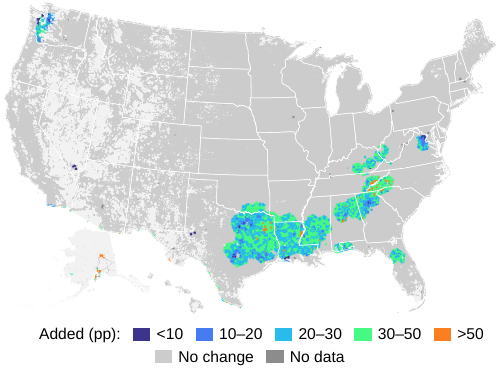}
    \caption{Additional share of households in each H3 cell priced out by constrained pricing (Max-only or demand surcharge), in percentage points relative to baseline prices; grey cells see no change.}
    \label{fig:afford_added}
  \end{subfigure}
  \caption{Starlink affordability. Unaffordability at baseline prices is widespread (left). Constrained pricing (right) makes Starlink unaffordable for 4.3M additional households compared to a baseline in which all cells have unconstrained prices.}
  \label{fig:afford}
\end{figure*}

Even without knowing why prices vary, we can still use this snapshot to understand what this variable pricing means for broadband affordability.
Here, we extend our analysis from our prior work~\cite{woo2025starlink} to define whether service is affordable.
We compare offered plans to household incomes, then check if the lowest priced Starlink plan would cost less than 2\% of a household's monthly income.
We adopt this 2\% threshold based on commonly used guidelines for Internet affordability~\cite{broadbandcommission2025affordability,fcc2016affordability,pew2023broadbandmiddleclass}.

To determine the number of households that are unable to afford Starlink service, we assign each location in the National Broadband Map (the public fabric~\cite{spin_fabric}) Starlink's lowest price from its H3 cell and the income distribution of its census tract (ACS 2020--2024 5-year estimates~\cite{acs5yr}, which count households in 16 income brackets).
Each location then ``counts'' as the share of its tract's households that earn too little to afford that price, and adding these up gives an estimate of how many households are priced out in each tract.
We amortize demand surcharges, equipment charges, and shipping over 12 months to compute an effective first-year monthly cost per plan, and use the cheapest plan available at each location.
We use the prices Starlink shows at checkout, including any promotions it applies automatically.\footnote{Starlink may change its prices, including promotions, with 30 days' notice~\cite[\S6.1]{starlink_tos}. At list prices, without promotions, 34\% of households (36 million) would be priced out, and constrained pricing would add 2.6 million.}
In particular, a promotion waived the \$55/mo plan's \$349 equipment charge at checkout in every cell where we observed that plan's cart, so the lowest-cost plan Starlink offers effectively costs about \$57/mo in its first year, including shipping.\footnote{We note that, should this free equipment promotion end, the effective cost of the lowest cost plan increases to approximately \$85/mo, equivalent to the mid-range service plan that includes free equipment even without promotional pricing.}

Nationally, we estimate that Starlink is not affordable for nearly a quarter of households (roughly 25.5 million; Figure~\ref{fig:afford}).
Most of this is driven by the base price:
even if every H3 cell were unconstrained, 20\% would be priced out, because a first-year cost of about \$57/mo translates to an annual household income of around \$34k to stay under 2\%.
In constrained areas, Starlink would be unaffordable for about half of households (49\%), compared to 20\% if all of these locations were unconstrained: \textbf{effectively, increased pricing due to constrained capacity makes Starlink unaffordable to an additional 4.3 million households today.}
Among un- and underserved households (i.e., those without terrestrial service providing 100/20 speeds), the gap is similar: 57\% of households in constrained areas are priced out, compared to 25\% if these locations were unconstrained.

\subsection{Implications for BEAD}
\label{ss:bead}

We next consider the implications for BEAD~\cite{bead}.
As part of NTIA's changes to BEAD eligibility in the so-called ``Benefit of the Bargain'' round in 2025~\cite{ntia2025beadrestructuring}, Starlink and other LEO satellite Internet providers were given the opportunity to bid on locations for BEAD awards, despite Starlink already claiming to provide 100/20 service to nearly the whole country.
The justification was that BEAD funds would pay LEO providers to reserve enough capacity on their networks to serve each funded location.

Unlike other ISPs that receive BEAD funds to expand their network to serve new unserved locations, Starlink already claims to serve the locations in which they've received BEAD awards, since they claim service at nearly every location in the United States.
Our dataset offers a new angle to consider in investigating where Starlink is currently unable to meet the demand.
Of the locations where Starlink is slated to receive BEAD awards as of October 2026~\cite{fcc_funding_map,bbexpanded2026bead}, 1 in 5 (about 90,000, largely in Texas, Washington, Louisiana, and Georgia) are in constrained H3 cells, and about 24,000 of those face a demand surcharge today.
NTIA's rules for satellite awards (``LEO Capacity Subgrants''~\cite{ntia2025beadrestructuring}) require providers to reserve capacity for every funded location and to offer service ``with no charges or delays attributable to extension of the service.''
Today's surcharges in these approximately 24,000 locations would seem to contradict that requirement.
With that said, BEAD awards for LEO satellite providers are still in flux, and these requirements only take effect once the provider certifies they are providing BEAD service, which Starlink has not done. 
We treat this as a baseline to revisit once these awards are finalized.

Nevertheless, 80\% of locations that Starlink is currently slated to receive funds for are in unconstrained H3 cells.
Simply offering unconstrained prices does not necessarily mean that Starlink doesn't need support to expand the capacity of its network, nor can we rule out the possibility that Starlink has already added capacity to their network in expectation of receiving these awards, thus clearing constrained areas that previously existed.
However, we argue it raises a question of what the marginal social value of subsidizing Starlink's capacity with public broadband funding is for locations they already claim to serve and in which they do not appear to face capacity constraints today.

\section{Conclusion}
\label{s:conclusion}
In this paper, we conduct the first national-scale evaluation of Starlink pricing.
We identify the spatial unit within which Starlink's prices and offers are uniform, the H3 resolution 6 grid cell, and find that Starlink varies prices both by adding explicit demand surcharges and by limiting the availability of lower-cost plans.
These constrained prices are offered in 9.8\% of H3 cells throughout the United States.
Using an affordability threshold of 2\% of monthly income, we find that Starlink as priced is not affordable for approximately 25.5 million households; constrained prices are responsible for making the service unaffordable for 4.3 million households.
Finally, we find that approximately 80\% of locations in which Starlink was slated to receive BEAD funding as of October 2026 have service available at unconstrained prices, raising the question of what additional service public funding secures at these locations.
These results highlight a gap between where Starlink is available and where it is affordable: the FCC's National Broadband Map and Starlink's own service availability map do not capture service costs or restrictions on offerings, both of which determine whether a household can actually adopt service.

We conclude by noting that understanding how prices change over time is at least as important as any single snapshot.
Soon after our survey was completed, Starlink began moving its residential service in the United States to ``dynamic pricing''~\cite{techspot2026dynamic}, under which prices may change more frequently and vary more finely across locations.
Meanwhile, the FCC no longer requires ISPs to publish machine-readable Broadband Consumer Labels~\cite{fcc2026bcl}, removing one of the few public sources of pricing data.
As prices change more often and become harder to observe, repeated, independent, national-scale measurement assumes an important role in reliably tracking broadband affordability and resulting gaps in digital inclusion.

\begin{acks}
We are grateful to Bright Data for their support of this project via the Bright Initiative.
This material is based upon work supported by the NSF under Award \#2542894.
Any opinions, findings and conclusions or recommendations expressed in this material are those of the authors and do not necessarily reflect the views of the NSF.
We additionally thank Wesley Woo for raising the questions that led to this study, as well as Robert Martin for his thoughtful comments on an earlier draft of this work.

\end{acks}

\bibliographystyle{ACM-Reference-Format}
\bibliography{refs}

\end{document}